\documentstyle[12pt]{article}

\input epsf

\catcode`\@=11
\@addtoreset{equation}{section}

\catcode`\@=12

\newtheorem{theor}{Theorem}

\newcommand{\om}{\omega}
\newcommand{\Om}{\Omega}
\newcommand{\pa}{\partial}
\newcommand{\ve}{\varepsilon}
\newcommand{\bR}{\bf R}
\newcommand{\const}{\mathop{\rm const}}
\newcommand{\codim}{\mathop{\rm codim}}
\newcommand{\sing}{\mathop{\rm sing}}
\newcommand{\supp}{\mathop{\rm supp}}
\renewcommand{\div}{\mathop{\rm div}}
\newcommand{\ol}{\overline{l}}
\newcommand{\oom}{\overline{\omega}}
\newcommand{\of}{\overline{f}}
\newcommand{\oC}{\overline{C}}

\newcommand{\oq}{\overline{q}}
\newcommand{\cA}{{\cal A}}
\newcommand{\cB}{{\cal B}}

\newcommand{\cT}{{\cal T}}
\newcommand{\cR}{{\cal R}}
\newcommand{\cF}{{\cal F}}
\newcommand{\cM}{{\cal M}}
\newcommand{\cL}{{\cal L}}
\newcommand{\cP}{{\cal P}}
\newcommand{\od}{\stackrel{\rm def}{=}}

\begin{document}

\title{Weak singularity dynamics in a nonlinear viscous medium}
\author{\thanks{This research was partially supported by the
Russian Foundation for Basic Research under grant No.~99-01-01074}}
\author{G.~A.~Omel'yanov\thanks{Moscow 
State Institute of Electronics and Mathematics,
B.Trekhsvyatitel'skii per. 3/12, Moscow 109028, Russia,
e-mail: pm@miem.edu.ru}}

\date{}
\maketitle

\begin{abstract}
We consider a system of nonlinear equations which can
be reduced to a degenerate parabolic equation. 
In the case $x\in\bR^2$ we obtained necessary conditions for the
existence of a weakly singular solution of heat wave type 
($\codim\sing\supp=1$) and of vortex type ($\codim\sing\supp=2$).
These conditions have the form of a sequence of differential
equations and allow one to calculate the dynamics of the
singularity support.
In contrast to the methods used traditionally for degenerate
parabolic equations, our approach is not based on comparison
theorems. 


\end{abstract}

\section{Introduction}

We consider the system of equations arising in problems 
of water purification:
\begin{eqnarray}
\frac{\pa c}{\pa t}&=&D\Delta c-D\kappa\div
\big(c\nabla(\div u-3\beta c)\big),\\
\Delta u&=&(3\omega-1)\beta\nabla c,
\end{eqnarray}
which describes diffusion in media with nonlinear viscosity.
Here $c\geq0$ is a scalar function, $u$ is a vector,
$x\in\Omega\subset\bR^n$, $t>0$, $D,\kappa,\beta>0$ and $\omega>1$
are several constants. 

We understand all derivatives in (1.1), (1.2) in the weak sense,
consider the case $n=2$, and obtain necessary conditions
for the existence of a solution of Eqs.~(1.1) and~(1.2) with a weak
singularity. These conditions, which allow us to calculate the
dynamics of a singularity, are obtained by using the asymptotic
expansion of the solution with respect to smoothness.

Prior to studying (1.1), (1.2), we consider the model example 
\begin{equation}%3
\frac{\pa u}{\pa t}=\frac12\frac{\pa^2}{\pa x^2}u^2,\qquad x\in\bR^1.
\end{equation}
One can readily see that a necessary and sufficient
condition for the existence of a solution  finite in~$x$ 
is satisfied for (1.3) (e.g., see~[1, 24] and the references
therein). 
On the other hand, from the viewpoint of distributions,
only the elements of $D'$ that admit multiplication 
can be singular solutions of a nonlinear equation.
As is well known~[18], 
in the one-dimensional case this restriction leads to an algebra
with generators~$1$, $\theta(x)$, $\ve\delta(x)$ 
and $x^\lambda_\pm$, $\lambda>0$.  
Here~$1$ stands for all smooth functions,  
$\ve\delta(0)=1$ and $\ve\delta(x)=0$ for $x\ne0$,
$$
\theta(x)=\begin{array}{cc}0,&x<0\\1,&x>0,\end{array}
\qquad
x^\lambda_+=\begin{array}{cc}0,&x\leq0\\x^\lambda,&x\geq0,\end{array}
\qquad
x^\lambda_-=\begin{array}{cc}|x|^\lambda,&x\leq0\\0,&x\geq0,\end{array}
$$
and $\theta^2=\theta$, $(x^\lambda_\pm)^2=x^{2\lambda}_\pm$, and
$\theta(x) x^\lambda_+=x^\lambda_+$. 

One can readily verify that both the Heaviside function 
and $\ve\delta$ cannot satisfy Eq.~(1.3). Nevertheless, if we set
$u=\big(x-a(t)\big)^\lambda_\pm$, then this singularity is
admissible for  $\lambda=1$.  
Such solutions of  degenerate parabolic equations are
well known and are called the {\it heat wave}.
In particular,
the simplest exact solution of Eq.~(1.3) is the one-parameter
family (e.g., see [24]):
\begin{equation}%4
u=(6t)^{-1}(x+\eta t^{1/3})_+(x-\eta t^{1/3})_-,\qquad 
\eta=\const>0,\quad t>0.
\end{equation}

A natural multidimensional analog of the heat wave 
$\big(x-a(t)\big)_+$ is the function $S_+$: 
$S_+=S$ for $S\geq0$ and $S_+=0$ for $S\leq0$, where $S=S(x,t)$
is a nondegenerate smooth function.
The fact that~$S$ is nondegenerate means that 
$\nabla S\big|_{S=0}\ne0$ for any chosen~$t$, and thus the
surface $\Gamma_t=\{x,S(x,t)=0\}$ is of codimension~$1$.  
Solutions of this type for a wide class of  degenerate
equations have been studied thoroughly (e.g., see [1, 7, 10, 24]). 
In particular, it is known that, 
starting from some time, 
the degeneration surface $\Gamma_t$ becomes differentiable 
[1, 5, 13, 22]. 

Another possible generalization of $\big(x-a(t)\big)_+$ 
to the multidimensional case is a singularity
of the form $|x-a(t)|$, $x,a\in\bR^n$.
Solutions of such form, whose singular support is of
codimensions $>1$, 
are called  {\it vortex type\/} solutions, since
they are used for the mathematical description of 
a typhoon eye motion [4]. 
It is known that  vortex type solutions exists for shallow
water equations [4]. 
However, as far as we know, the existence problem for such
solutions of parabolic equations has not been discussed. 

In this paper we obtain necessary conditions under which 
system~(1.1), (1.2) have  singular solutions of both types mentioned
above. 

We study system (1.1), (1.2) by using the asymptotic expansion of the
solution with respect to smoothness. For linear equations, the
idea to expand the solution with respect to smoothness was
proposed and developed by R.~Courant, D.~Ludwig, V.~P.~Maslov, 
V.~M.~Babich and others.
A further development of this idea applied to quasilinear
equations and special solutions lying in $D'$ and admitting
multiplication was performed by many authors [7, 8, 23] 
starting from Maslov's paper [19].
A possibility to study a more general class of nonsmooth
solutions of quasilinear equations [2, 3, 6, 9, 11, 12, 21] appeared after the
construction of algebras of generalized functions, 
which include distributions from~$D'$ 
(V.~K.~Ivanov, G.~Colombeau, Yu.~V.~Egorov, and
others). 

A method based on asymptotic expansions of solutions with respect
to smoothness is very fruitful, 
since in this method 
the maximum principle and the comparison theorems 
are not supposed to be valid. On the other hand, 
the following significant drawback is typical: 
this method implies a system of model equations 
(the so-called {\it Hugoniot type conditions\/}),
which is an infinite nontriangular sequence 
in the case of nonlinear equations.

For example, by setting 
\begin{equation}%5
u=a_1(t)(x-\varphi)_++a_2(t)(x-\varphi)^2_+ + \dots
\end{equation}
in a neighborhood of the singularity support $x=\varphi(t)$ and
substituting (1.5) into (1.3), we obtain the system 
\begin{equation}%6
\dot\varphi=-a_1,\qquad 
\dot a_1=4a_1a_2,\qquad
\dot a_2=3(2a^2_2+3a_1a_3),\dots
\end{equation}
In the special case $a_1(t_0)=\eta/(3t_0^{2/3})$,
$a_2(t_0)=-1/(6t_0)$, $a_i(t_0)=0$, $i\geq3$, 
for $t\geq t_0>0$ it follows from ~(1.6) that 
$$
\varphi=-\eta t^{1/3},\qquad a_1=\eta/3t^{2/3},\qquad
a_2=-1/6t,\qquad a_i=0,\qquad i\geq3.
$$
Thus we arrive at the exact solution~(1.4) considered near the
left-hand front $x=-\eta t^{1/3}$.

In the general case system~(1.6) is an infinite sequence of
coupled equations. Of course, infinite sequences cannot be used
for practical computations, hence we need to close them. 
If $\codim\sing\supp u=1 $, then this can be performed by
passing to a free boundary problem. In this case  conditions of
the free boundary follow from a local description of the
solution near the singularity support and from the first
Hugoniot type condition. 
For example, 
it follows from this procedure that  in
the region $Q_t=\Om_t\times\{t>t_0\}$, 
$\Omega_t=\{x,\varphi_-(t)<x<\varphi_+(t)\}$
we need to study Eq.~(1.3) with the conditions
\begin{equation}%7
u\bigg|_{x=\varphi_\pm\pm0}=u\bigg|_{x=\varphi_{\pm}\mp0}=0,\qquad 
\dot \varphi_{\pm}=\frac{\pa u}{\pa\nu}\bigg|_{x=\varphi_\pm\mp0}
\end{equation}
and a natural initial condition. 
The first condition in~(1.7), which follows from the structure of
solution~(1.5), implies that ~$u$ and the flow $uu_x$ are
continuous on $\pa\Om_t$.
The second condition in (1.7), which relates the boundary velocity 
to the limit value of the derivative of~$u$ along the outward
normal~$\nu$, 
is exactly the first Hugoniot type condition in~(1.6). 
In this statement of the problem we can ignore all other
conditions in~(1.6), since they are automatically satisfied due
to the fact that the free boundary problem has a solution. 

From the ideological viewpoint, 
this method of closing system~(1.6) is similar 
to closing the Hugoniot type conditions for the shock wave 
by passing to the classical statement of the problem describing
the shock wave  motion ~[8]. 
However, here we obtain a problem that differs from the
classical one-phase Stefan problem 
(see [20] and [14, 25] for related problems)
by degeneration of the parabolic operator on the free
boundary. 

Another possible approach is to truncate the sequence of
Hugoniot type conditions. 
By choosing a number $N>0$ and setting all ``extra'' function
equal to zero, we can calculate the dynamics of a singularity by
using the first~$N$ Hugoniot type conditions. 
However, in this cases we need to justify this procedure. 
In general, this problem, which is similar to the well-known
problem of truncating Bogolyubov--Born--Green-Kirkwood--Yvon
chains, remains open.  
Nevertheless, the first result in justifying the truncation
procedure has been obtained [8],  and numerical results show
a good agreement between the exact solution and those obtained by
using truncated sequences [4, 23] for a reasonably large time
interval.  

In the case $\codim\sing\supp u>1$ the truncation procedure is
apparently the only real method for calculating the singularity
dynamics. 
Indeed, in this case, instead of the free boundary problem, we
must consider a nonlinear equation in a domain with a remote set
of codimension $>1$. 
The position of this set, on which we pose additional
conditions, varies in time and is to be determined.
As far as we know, such problems have not been studied till
recently.  

Now let us consider system (1.1), (1.2) and write it in a simple
form. 
To this end, we apply (in the $D'$-sense) the operator $\div$ to
Eq.~(1.2). We assume that~$c$ has a singularity of an above-listed
type and obtain the relation 
$$
\Delta\big(\div u-\beta(3\omega-1)c\big)=0,
$$
which holds for all~$x$ including the singularity points. Hence 
$$
\div u=\beta(3\omega-1)c+w,
$$
where $w$ is an arbitrary harmonic function. 
Now Eq.~(1.1) acquires the form 
\begin{equation}%8
\frac{\pa c}{\pa t}=D\big\{\div\big((1-\kappa\beta
(3\omega-4)c)\nabla c\big)-\kappa\langle\nabla c,\nabla w\rangle\big\}.
\end{equation}
We assume that $\omega>4/3$, since only in this case Eq.~(1.8) is
degenerate and, correspondingly, can have singular solutions
bounded in the $C$-norm.

By changing the variables, we rewrite (1.8) in the compact form
\begin{equation}%9
\frac{\pa c}{\pa t}+\langle\nabla w,\nabla c\rangle
=\Delta(c-\mu c^2),\qquad \mu=\const>0.
\end{equation}

Our goal is to derive necessary conditions for the existence of
a solution with a special local structure. Therefore, we abstract
our consideration from the initial and boundary (on the outer
boundary $\pa\Om$) conditions. 
However, it is clear that in this case the initial data cannot
be arbitrary but must satisfy requirement that follow from the
Hugoniot type conditions obtained below.

In the present paper we restrict ourselves to a discussion of
Eq.~(1.9). However, the method proposed here for constructing
singular solutions can be easily generalized to a wide class of
degenerate parabolic equations and systems.

\section{Heat wave dynamics}

In~$D'$ it readily follows from Eq.~(1.9) that for all~$t$ 
the wave front  $\Gamma_t=\{x,S(x,t)=0\}$ lies on the
degeneration surface $\Sigma=\{(x,y),c=1/2\mu\}$ corresponding
to Eq.~(1.9). This fact implies that the asymptotic expansion of
the solution with respect to smoothness in the neighborhood of
$\Gamma_t$ has the form
\begin{equation}%10
c=(2\mu)^{-1}+a_1S_+ +a_2S^2_+ +\dots
\end{equation}
Here 
$S\in C^\infty$ is a function such that 
$\nabla S\big|_{\Gamma_t}\ne0$, $a_1<0$,
the coefficients $a_i=a_i(x,t)\big|_{\Gamma_t}\in C^\infty$ have
the meaning of the normal derivatives of~$c$ on~$\Gamma_t$, and 
$\pa a_i/\pa\nu\big|_{\Gamma_t}=0$, 
where 
$\nu=-\nabla S/|\nabla S|\big|_{\Gamma_t}$ is the outward
normal. Just as in the case of the model equation (1.3), 
the structure of the solution guarantees that 
the flow $(1-2\mu c)\pa c/\pa\nu$ is continuous on $\Gamma_t$.
To find~$S$ and~$a_i$, we substitute (2.1) into (1.9) and expand 
$\nabla w$ in the Taylor series on $\Gamma_{t}$:
$$
\nabla w=V_0+V_1 S+V_2 S^2+\dots
$$

Now by setting the coefficients of $\theta(S), S_+,S^2_+,\dots$  
equal to zero, we obtain the infinite nontriangular system 
\begin{eqnarray}%11
&&\{S_t+\langle V_0,\nabla S\rangle+2\mu|\nabla S|^2a_1\}\Big|_{S=0}=0,\\
&&\{a_{i_t}+\langle V_0,\nabla\psi\rangle a_{i\xi}
+2(i+1)^2\mu|\nabla S|^2a_1a_{i+1}+F_i\}\Big|_{S=0}=0,\quad i=1,2,\dots
\nonumber
\end{eqnarray}
Here $\xi=\psi(x,t)$ is a parameter along $\Gamma_t$ that is
determined in each chart as the solution of the equation $S=0$,
and $F_i$ are functions of $S,a_1,\dots,a_i$ and their derivatives. 
In particular, we have 
\begin{eqnarray*}
F_1&=& a_1\langle V_1,\nabla S\rangle +2\mu a^2_1\Delta S,\\
F_2&=& 2a_2\langle V_1,\nabla S\rangle 
+a_{1\xi}\langle V_1,\nabla\psi\rangle
+a_1\langle V_2,\nabla S\rangle\\
&&
+\mu\{6a_1a_2\Delta S+12 a^2_2|\nabla S|^2
+|\nabla\psi|^2(a^2_1)_{\xi\xi}+\Delta\psi(a^2_1)_\xi\}.
\end{eqnarray*}

Equations (2.2) are written in the form that is inconvenient for
analysis and numerical calculations, since in (2.2) we need to
calculate the trace on $\Gamma_t$. This can be avoided by taking
into account the geometry of the front $\Gamma_t$. 
For example, if for $t\in[0,T)$ the front can be uniquely
projected on the $x_2$-axis, then we can set 
$S=x_1-\varphi(x_2,t)$,  
obtain $\nabla S=(1,-\varphi_{x_2})^T$, $\xi=x_2$, 
$a_i=a_i(x_2,t)$, and rewrite Eq.~(2.2) as 
\begin{eqnarray}%12
&&\varphi_t+w_{x_2}\Big|_{x_1=\varphi}\varphi_{x_2}
=2\mu(1+\varphi^2_{x_2})a_1+w_{x_1}\Big|_{x_1=\varphi},\\
&&a_{1_t}+w_{x_2}\Big|_{x_1=\varphi}a_{1_{x_2}}
+a_1\langle V_1\Big|_{x_1=\varphi},\nabla S\rangle\nonumber\\
&&\qquad
+2\mu\big(-a^2_1\varphi_{x_2x_2}-2(a^2_1)_{x_2}\varphi_{x_2}
+4(1+\varphi^2_{x_2})a_1a_2\big)=0,\dots,
\nonumber
\end{eqnarray}

If $\Gamma_t$ is a closed smooth curve without
self-intersections, then we can avoid a local description of the
front by setting $S=t-\Phi(x)$. 
Then $a_i=a_i(x)$ and, instead of (2.2), we obtain
\begin{eqnarray}%13
&&1-\langle\nabla\Phi,\nabla w\rangle\Big|_{t=\Phi}
+2\mu a_1|\nabla\Phi|^2=0,\\
&&-\langle\nabla a_1,\nabla w\rangle\Big|_{t=\Phi}
+a_1\langle V_1,\nabla\Phi\rangle\nonumber\\
&&\qquad
+2\mu a_1\{4\langle\nabla\Phi,\nabla a_1\rangle
+a_1\Delta\Phi-4a_2|\nabla\Phi|^2\}=0,\dots,\nonumber
\end{eqnarray}
where $V_1=(\nabla w_t)\big|_{t=\Phi}$.

By writing the Hugoniot conditions in the form (2.3) or (2.4), we
arrive at more trivial problem of posing the initial conditions.
Namely, for~(2.3) we specify the values of $\varphi_1,a_1,\dots$
for $t=0$,  and for~(2.4) we specify the values $\Phi=0$ and 
$a_1=a^0_1,\dots$ on the initial curve~$\Gamma_0$ whose position
is assumed to be known in advance.  

Thus we arrive at the statement.

\begin{theor}%Theorem~1. 
For the existence
of a heat wave type solution of Eq.~{\rm(1.9)}, 
is necessary that the Hugoniot type conditions {\rm(2.2)} be
satisfied on the front $\Gamma_t$.
\end{theor}

Note  that $a_1|\nabla S|\big|_{\Gamma_t}
=-\pa c/\pa \nu\big|_{\Gamma_t}$ and $v_\nu=S_t/|\nabla S|$ is
the velocity of the boundary $\Gamma_t$ along its outward
normal. Therefore, the first Hugoniot type condition can be
written in the form 
\begin{equation}%14
v_\nu=2\mu\frac{\pa c}{\pa\nu}\bigg|_{\Gamma_t}
-\frac{\pa w}{\pa\nu}\bigg|_{\Gamma_t}.
\end{equation}

As was already noted in Introduction, condition (2.5), together
with the condition $c\big|_{\Gamma_t}=(2\mu)^{-1}$, allows us
to reduce calculations of the heat wave to a one-phase free
boundary problem. 
Obviously, the solvability of this problem implies that 
all Hugoniot type conditions are satisfied. 

\medskip

\noindent
{\bf Remark}
If instead of (1.9) we consider the equation 
$$
\frac{\pa c}{\pa t}+\langle V,\nabla c\rangle=
\langle \nabla,k(c)\nabla c\rangle,
$$
where $k=k_0c^\gamma(1+O(c))$ as $c\to0$, $k_0>0$, and
$\gamma>0$, 
then by analogy with (2.1) the asymptotic expansion of the
solution with respect to smoothness has the form
$$
c=a_1S^{\gamma^{-1}}_+ +
a_2S^{\gamma^{-1}+1}_+ +\dots
$$
In this case the first Hugoniot type condition is similar to
(2.2), namely, 
$$
\Big\{S_t+\langle V,\nabla S\rangle
-\frac{k_0}{\gamma}|\nabla S|^2a^\gamma_1\Big\}\Big|_{S=0}=0.
$$

However, now the closure of the sequence of Hugoniot type
conditions  leads to a one-phase Stefan type problem, which
consists of the original equation considered in the domain 
$Q_t=\Omega_t\times\{t>t_0\}$,
$\Omega_t=\{x,S(x,t)>0\}$ with the initial condition and the
boundary conditions on the free boundary $\Gamma_t=\pa\Omega_t$:
$$
c\Big|_{\Gamma_t}=0,\qquad 
v_\nu=\langle V,\nu\rangle\Big|_{\Gamma_t}
-k_0c^{\gamma-1}\frac{\pa c}{\pa \nu}\Big|_{\Gamma_t},
$$
where $f\big|_{\Gamma_t} $ is understood, just as in (2.2), as
the limit obtained by passing to $\Gamma_t$ along the inward normal. 
Obviously, we can rewrite the last term in the Stefan condition
in the form $-k(c)c^{-1}\pa c/\pa\nu\big|_{\Gamma_t} $.

\medskip

\section{Dynamics of a vortex type singularity}

The solution of Eq.~(1.9) with a vortex type singularity can be
written in the form
\begin{equation}%15
c=c^0(x,t)+\sqrt{S(x,t)}c^1(x,t),
\end{equation}
where $S,c^0,c^1$ are some smooth function.
We assume that for each fixed $t$
\begin{equation}%16
S\geq0,  \qquad 
\nabla S\Big|_{S=0}=0,\qquad 
S''\Big|_{S=0}>0,  
\end{equation}
and moreover, 
$\cT=\bigcup_t\Gamma_t\subset \bR_+\times \bR^2$ is a curve 
in the $C^\infty$-class. 
Here $\Gamma_t=\{x,S(x,t)=0\}$ is the singular support of the
solution and $S''$ is the Hessian of~$S$. 
By $a(t)$ we denote a vector-function, which describes the
position of a singularity at each time instant~$t$, i.e., 
we have $\Gamma_t=\{x,x+a(t)=0\}$.
We perform the change $x'=x+a(t)$. Then, omitting the prime
on the new variable, instead of Eq.~(1.9), we obtain the equation 
\begin{equation}%17
\dot c+\langle\dot a+\nabla w,\nabla c\rangle=\Delta(c-\mu c^2).
\end{equation}
Here and in the following, the dot indicates the derivative with
respect to~$t$.

Substituting (3.1) into (3.3) and grouping smooth and singular
terms, we obtain the relation 
\begin{equation}%18
D^0+S^{-3/2}D^1=0,
\end{equation}
where 
\begin{eqnarray}%19
D^0&=& \dot c^0+\langle V,\nabla c^0\rangle-\Delta G^0,\qquad 
V=\dot a+\nabla w,\nonumber\\
D^1&=&-\kappa G^1+4S\Big(\frac12\langle \nabla S,V\rangle c^1
-\langle\nabla S,\nabla G^1\rangle\Big)\\
&&{}
+4S^2(-\Delta G^1+\dot c^1+\langle V,\nabla c^1\rangle), \qquad 
G^0=c^0-\mu({c^0}^2+S{c^1}^2),\nonumber\\
G^1&=&(1-2\mu c^0)c^1,\qquad 
\kappa=2S\Delta S-|\nabla S|^2. \nonumber
\end{eqnarray}
The smooth functions $D^i$ satisfy relation (3.4) at the point $x=0$
only if for any $N\geq0$ we have
\begin{eqnarray}%20,21
\frac{\pa^k}{\pa x^\alpha}D^0\bigg|_{x=0}&=&0,\qquad 
|\alpha|=k,\qquad k\leq N,\\
\frac{\pa^k}{\pa x^\alpha}D^1\bigg|_{x=0}&=&0,\qquad 
|\alpha|=k,\qquad k\leq N,
\end{eqnarray}
where, as usual, $\alpha=(\alpha_1,\alpha_2)$ is a multiindex.

We write the functions $S$ and $c^i$, $i=0,1$, as the formal
Taylor expansions
\begin{eqnarray}%22
c^i&=&\sum^\infty_{k=0}p^i_k(x,t),\qquad 
S=\sum^\infty_{k=2}S_k(x,t),\\
p^i_k&=&\sum_{|\alpha|=k}c^i_\alpha(t)x^\alpha,\qquad 
S_k=\sum_{|\alpha|=k}s_\alpha(t)x^\alpha.\nonumber
\end{eqnarray}
In view of conditions (3.2), we assume that $s_{11}=0$, 
$s_{20}>0$, and $s_{02}>0$.

In what follows, we essentially use the fact that the
vector-function~$V$ is of a special structure. Namely, 
since $w$ is a harmonic function, we have
\begin{equation}%23
w=\sum^\infty_{k=0}W_k(x,t),\qquad 
W_k=\sum_{|\alpha|=k}\omega_\alpha(t)x^\alpha,
\end{equation}
where $W_k$ are harmonic polynomials. Therefore, in particular, 
$W_0=W_0(t)$ and $W_1(x,t)$ are arbitrary functions, and  
\begin{eqnarray}%24
W_2&=&\omega_{20}(x^2_1-x^2_2)+\omega_{11}x_1x_2,\\
W_3&=&\omega_{30}x_1(x^2_1-3x^2_2)+\omega_{03}x_2(x^2_2-3x^2_1),
\nonumber
\end{eqnarray}
where $\omega_\alpha=\omega_\alpha(t)$ are arbitrary smooth
functions. Thus in the Taylor expansion 
\begin{equation}%25
V=\sum^\infty_{k=0}V_k(x,t),\qquad 
V_k=\sum_{|\alpha|=k}v_\alpha(t)x^\alpha,
\end{equation}
we must take into account that $V_0=\dot a+\nabla W_1$ and 
$V_i=\nabla W_{i+1}$, $i\geq1$.

Substituting expansions (3.8)--(3.11) into (3.6) and (3.7) and
setting the coefficients of $x^\alpha$ equal to zero, we obtain
an infinite system of equations. 
One can readily see that this system is not triangular, i.e.,
the first~$N$ equations contain $M>N$ unknowns. 
The inequality $M\geq N$ is necessary for the existence 
(at least, for small~$t$) of a solution of these equations, but,
in general, is not sufficient. 
The matter is that the compatibility conditions for these
equations contain less than~$M$ unknown functions. 
Since we do not know in advance whether the equations obtained
are solvable, a detailed analysis of these equations is
required.  

To avoid too cumbersome formulas, we shall successively analyze
the equations obtained for $k=0,1,\dots$

We consider (3.7) for $k=2$. Since 
\begin{equation}%26
\kappa=\sum^\infty_{k=2}\kappa_k(x,t),\qquad 
\kappa_2=2S_2\Delta S_2-|\nabla S_2|^2=4s_{20}s_{02}x^2,
\end{equation}
we 
have the relation
$$
s_{20}s_{02}c^1_0(1-2\mu c^0_0)=0.
$$
In view of (3.2) and the natural assumptions that 
\begin{equation}%27
c^1_0\ne0
\end{equation}
(otherwise, the singularity of the solution (3.1) becomes weaker
that $|x|$), 
we derive the relation
\begin{equation}%28
c^0_0=1/2\mu,
\end{equation}
which readily implies that the condition $\mu>0$ is necessary for 
the existence of a physically meaningful ($c\geq0$) solution of
the form~(3.1). 

Now we consider (3.7) for $k=3$. Since $G^1_0=0$ in view of (3.14),
we have the relation
\begin{equation}%29
\kappa_2 G^1_1+4S_2\langle \nabla S_2,\nabla G^1_1-\frac12 V_0 c^1_0\rangle=0.
\end{equation}
Here and in the following, relations of the form 
$P_k(x,t)\equiv \sum_{|\alpha|=k}P_\alpha(t)x^\alpha=0$ mean
that the coefficients $P_\alpha(t)$, $|\alpha|=k$, are equal to
zero.  
We also use expansions of the functions~$G^i$
that are similar to (3.8):
\begin{equation}%30
G^i=\sum^\infty_{k=0}G^i_k(x,t),\qquad 
G^i_k=\sum_{|\alpha|=k}g_\alpha(t)x^\alpha,\quad i=1,2.
\end{equation}
In particular, by taking into account (3.14), we have
\begin{eqnarray}%31
G^0_2&=&-\mu(2c^1_0p^1_2+{p^0_1}^2+{p^1_1}^2),\qquad G^1_0=0,\\
G^1_1&=&-2\mu c^1_0p^0_1,\qquad G^1_2=-2\mu(c^1_0p^0_2+p^1_1p^0_1).\nonumber
\end{eqnarray}
By substituting explicit formulas for $\kappa_2$, $S_2$, $G^1_1$
into (3.15) and matching the  coefficients of like powers of~$x$,
we obtain the four 
equations 
\begin{equation}%32
6\mu c^0_\alpha+V_{0,\alpha}=0,\qquad
2\mu(s_{02}/s_{20})^{q_\alpha}c^0_\alpha+V_{0,\alpha}+4\mu c^0_\alpha=0,
\end{equation}
where $\alpha$ attains the values $(1,0)$ and $(0,1)$, 
$V_{0,\alpha}=V_{0,1}$ and $q_{\alpha}=1$ for $\alpha=(1,0)$,
$V_{0,\alpha}=V_{0,2}$ and $q_{\alpha}=-1$ for $\alpha=(0,1)$,
and $V_{0,i}$ are components of the vector 
$V_0=(V_{0,1},V_{0,2})$.

Obviously, Eqs.~(3.18) are compatible only for $s_{20}=s_{02}$. 
Without loss of generality, we obtain  
\begin{equation}%33
s_{20}=s_{02}=1.
\end{equation}
The (3.18) implies the relation 
\begin{equation}%34
V_0+6\mu\nabla p^0_1=0
\end{equation}
between the velocity $\dot a$ of the singularity and the
coefficients in expansions (3.8) and (3.9).

Next, from (3.6) with $k=0$ we obtain 
$$
\dot c^0_0+\langle V_0,\nabla p^0_1\rangle=\Delta G^0_2.
$$
In view of (3.14), (3.19), and (3.20), this immediately implies
\begin{equation}%35
{c^1_0}^2=|\nabla p^0_1|^2.
\end{equation}

Now let us consider (3.7) for $k=4$.
In view of (3.19), we have
\begin{equation}%36
x^2 R_2+S_3 L_1=0
\end{equation}
with homogeneous polynomials $R_2$ and $L_1$ of degree~2
and~1, respectively. 
One can readily see that the compatibility condition 
for the equations obtained from (3.22) is 
$$
\Lambda\ol_1=0,
$$
where $\Lambda$ is the $2\times2$ matrix with coefficients 
$\Lambda_{11}=-\Lambda_{22}=s_{21}-s_{03}$, 
$\Lambda_{12}=\Lambda_{21}=s_{30}-s_{12}$
and $\ol_1=(l_{10},l_{01})$, where $l_\alpha$ are 
coefficients in the polynomial $L_1$. 
The choice $\ol_1=0$ implies that the solution constructed is
trivial. Therefore, 
$\det \Lambda=0$, and hence, in turn,  
\begin{equation}%37
S_3=x^2Q_1(x,t),
\end{equation}
where $Q_1$ is a homogeneous polynomial of the first order.

By using (3.23) and the obvious identity  
\begin{equation}%38
\langle x,\nabla p_n\rangle=np_n,
\end{equation}
we transform (3.22) as follows:
\begin{equation}%39
5p^0_2+Q_1p^0_1+\frac2{c^1_0}p^1_1p^0_1
+\frac1\mu W_2=\frac12x^2
\Big(\langle\nabla Q_1,\nabla p^0_1\rangle+\frac1{\mu c^1_0}\theta_0\Big),
\end{equation}
where
$$
\theta_0=-\dot c^1_0+6\mu\langle\nabla p^0_1,\nabla p^1_1\rangle
+\Delta G^1_2.
$$
Relation (3.25) implies three equations for the coefficients in
expansions (3.8). We combine these equations and, in particular,
obtain the system
\begin{equation}%40
{\cal A}\nabla Q_1=-\frac{2}{c^1_0}{\cal A}\nabla p^1_1-f,
\end{equation}
where ${\cal A}$ is the $2\times2$ matrix 
with coefficients $\cA_{11}=\cA_{22}=c^0_{01}$, 
$\cA_{12}=-\cA_{21}=c^0_{10}$, $f=(f_1,-f_2)$, 
$f_1=5c^0_{11}+\omega_{11}/\mu$, and 
$f_2=5(c^0_{20}-c^0_{02})+2\omega_{20}/\mu$.

By using (3.21) and the assumption (3.13), we find from (3.26) the
coefficients in the polynomial $Q_1$:
\begin{equation}%41
\nabla Q_1=-\frac2{c^1_0}\nabla p^1_1-\psi_0,\qquad 
\psi_0=\cA^{-1}f.
\end{equation}
Now the third equation obtained from (3.25) can be reduced to the
form 
\begin{equation}%42
\dot c^1_0+\frac 92\mu c^1_0\Delta p^0_2=0.
\end{equation}

Next, we consider (3.6) for $k=1$ and arrive at the relation 
\begin{equation}%43
\dot p^0_1-2\mu\langle\nabla p^0_1,\nabla p^0_2\rangle
+\langle V_1,\nabla p^0_1\rangle+2\mu(p^0_1\Delta p^0_2-4\psi_1)=0,
\end{equation}
where $\psi_1$ is a homogeneous polynomial such that 
$\nabla\psi_1={c^1_0}^2\psi_0$.
One can readily see that the system of relations (3.21), (3.28),
(3.29) is consistent only if the compatibility condition
\begin{equation}%44
\langle\nabla p^0_1,\cB\nabla p^0_1\rangle=0
\end{equation}
is satisfied. Condition (3.30) can be readily obtained by
differentiating (3.21) with respect to $t$ and calculating 
$\nabla\dot p^0_1$ with the help of (3.29). In (3.30) 
$$
\cB=2{p^0_2}'' +\frac52 I\Delta p^0_2+8\of-\mu^{-1}W''_2,
$$
$I$ is the unit $2\times 2$ matrix, the double prime indicates 
the Hessian of a function, and $\of$ is the $2\times 2$ matrix
with coefficients 
$\of_{11}=-\of_{22}=f_2$ and $\of_{12}=\of_{21}=f_1$.

It follows from (3.30) that 
\begin{equation}%45
\cB\nabla p^0_1=7\sigma\nabla p^{0\perp}_1,
\end{equation}
where $f^\perp=Tf$ is the vector orthogonal to~$f$, 
$T$ is the rotation matrix, i.e., 
$T_{11}=T_{22}=0$ and $T_{12}=-T_{21}=-1$, and 
$\sigma=\sigma(t)$ is an arbitrary scalar. 

Let us transform (3.31). To this end, we note that 
$\cB$ can be rewritten in the form $\cB=7(\cB^0+\mu^{-1}W''_2)$, 
where $\cB^0$ is the matrix with coefficients  
$\cB^0_{11}=7c^0_{20}-5c^0_{02}$, 
$\cB^0_{22}=-5c^0_{20}+7c^0_{02}$, 
$\cB^0_{12}=\cB^0_{21}=6c^0_{11}$.

Moreover, the following relations hold:
$W''_2\nabla p^0_1=D\oom_2$ and 
$\cB^0\nabla p^0_1=\cB^1\oC_2+c_{02}r$, 
where $\oom_2=(\om_{20},\om_{11})^T$, 
$\oC_2=(c^0_{20},c^0_{11})^T$, and $r=(-5c^0_{10},7c^0_{01})^T$, 
$D$ and $\cB^1$ are the matrices with coefficients 
$D_{11}=2D_{22}=2c^0_{10}$,
$D_{21}=-2D_{12}=-2c^0_{01}$,
$\cB^1_{11}/7=\cB^1_{22}/6=c^0_{10}$, and
$\cB^1_{12}/6=-\cB^1_{21}/5=c^0_{01}$.
Therefore, we can rewrite (3.31) in the form 
\begin{equation}%46
D\oom_2=\mu(\sigma T-\cB^0)\nabla p^0_1
=\mu\{\sigma\nabla p^{0\perp}_1-\cB^1\oC_2-c^0_{02}r\}.
\end{equation}
Both matrices $D$ and $\cB^1$ are nondegenerate in view of the
assumption (3.13) 
($\det D=2{c^1_0}^2$ and $\det\cB^1=6(5{c^1_0}^2+2{c^0_{10}}^2)\od 6d$).
However, we must take into account that any condition imposed
on~$W_2$ implies restrictions on the boundary values of~$u$. 
So, avoiding the appearance of such restrictions and inverting
$\cB^1$, we obtain from~(3.30) the algebraic relations
\begin{eqnarray}%47
c^0_{20}&=&\frac1d
\{-2\sigma c^0_{10}c^0_{01}+c^0_{02}(5{c^{0}_{10}}^2+7{c^{0}_{01}}^2)
-2{c^{1}_0}^2\om_{20}/\mu\},\\
c^0_{11}&=&\frac1{6d}
\{-\sigma(5{c^{0}_{01}}^2-7{c^{0}_{10}}^2)
-24c^0_{02} c^0_{10}c^0_{01}+4c^0_{10}c^0_{01}\om_{20}/\mu\nonumber\\
&&\qquad 
-(7{c^{0}_{10}}^2+5{c^0_{01}}^2)\om_{11}/\mu\}.\nonumber
\end{eqnarray}
Obviously, 
since the higher-order terms of the expansion contain
differential equations for the coefficients of $p^0_2$, 
we must prove the validity of relations (3.33). 

First, we note that (3.32) allows us to simplify Eq.~(3.29). Namely,
we invert the matrix $D$, calculate $V_1$ and $\psi_1$, and,
according to~(3.21), set 
\begin{equation}%48
c^0_{10}=c^1_0\sin\varphi,\qquad c^0_{01}=c^1_0\cos\varphi,
\end{equation}
where $\varphi=\varphi(t)$ is a new unknown function. 
Then after some calculations we transform (3.29) to the final form
\begin{equation}%49
\dot \varphi+7\mu\sigma=0.
\end{equation}
To prove that (3.33) is consistent, 
we consider (3.6) for $k=2$, which leads to the equation 
\begin{equation}%50
\dot p^0_2+\langle V_0,\nabla p^0_3\rangle
+\langle V_1,\nabla p^0_2\rangle
+\langle V_2,\nabla p^0_1\rangle=\Delta G^0_4,
\end{equation}
and (3.7) for $k=5$, which leads to the relation 
\begin{equation}%51
6\mu c^1_0 S_4 p^0_1+x^2 \cR_3=0,
\end{equation}
where $\cR_3$ is a homogeneous polynomial of the third order. 
By analogy with (3.23), we can readily verify that (3.37) implies
the relation 
\begin{equation}%52
S_4=x^2 Q_2(x,t),
\end{equation}
where $Q_2$ is a homogeneous polynomial of the second order. 
Then (3.37) acquires the form 
\begin{equation}%53
-4c^1_0p^0_1Q_2+\cR_3+x^2\{c^1_0
(\langle\nabla Q_2,\nabla p^0_1\rangle-p^0_1\Delta Q_2)
-\mu^{-1}\dot p^1_1+\cL_1\}=0,
\end{equation}
where
\begin{eqnarray*}
\cR_3&=&-2\big\{4p^0_1p^1_2+7c^1_0p^0_3+7p^1_1p^0_2
+Q_1\big(p^0_1(5p^1_1+c^1_0Q_1)+11c^1_0p^0_2\big)\\
&&\qquad 
+\big(W_2(p^1_1+2Q_1c^1_0)+3W_3 c^1_0/2\big)/\mu\big\},\\
\cL_1&=&Q_1(c^1_0\langle\nabla Q_1,\nabla p^0_1\rangle+2\theta_0/\mu)
+|\nabla Q_1|^2c^1_0p^0_1/2\\
&&\qquad 
+\mu^{-1}\langle\nabla Q_1,\nabla G^1_2-(p^1_1V_0+c^1_0V_1)/2\rangle\\
&&\qquad 
+\mu^{-1}
(\Delta G^1_3
-\langle V_0,\nabla p^1_2\rangle
-\langle V_1,\nabla p^1_1\rangle).
\end{eqnarray*}
Relation (3.39) is equivalent to four scalar equations. The
compatibility condition for these equations is
$$
\cA\oq_2=- q_{02}\nabla p^{0\perp}_1-F/4c^1_0,
$$
where $\cA$ is the same matrix as in (3.26), 
$\oq_2=(q_{20},q_{11})^T$, 
$q_\alpha$, $|\alpha|=2$, are the coefficients in the
polynomial~$Q_2$, 
and $F$ is the vector with components
$F_1=r_{03}-r_{21}$ and $F_2=r_{30}-r_{12}$, where  $r_\alpha$
are coefficients in~$\cR_3$.  
Hence we find 
\begin{equation}%54
q_{20}=q_{02}+\langle F,\nabla p^{0\perp}_1\rangle/4{c^1_0}^3,
\qquad
q_{11}=-\langle F,\nabla p^{0}_1\rangle/4{c^1_0}^3,
\end{equation}
and see that $q_{02}$ is arbitrary.

Now it follows from (3.40) that the other two relations that
follow from (3.39) are
\begin{equation}%55
\nabla \dot p^1_1=\mu\cF-6\mu c^1_0 q_{02}\nabla p^0_1.
\end{equation}
Here $\cF$ is the vector with components
\begin{eqnarray*}
\cF_{1}&=&r_{30}+l^1_{10}
-\langle F,4c^0_{10}\nabla p^{0\perp}_1
+c^0_{01}\nabla p^0_1\rangle/4{c^1_0}^2,\\
\cF_{2}&=&r_{03}+l^1_{01}
-\langle F,2c^0_{01}\nabla p^{0\perp}_1
+c^0_{10}\nabla p^0_1\rangle/4{c^1_0}^2,
\end{eqnarray*}
where $l^1_\alpha$ are coefficients in $\cL_1$.

In turn, relations (3.40) allow us to rewrite Eq.~(3.36) in the form
\begin{equation}%56
\dot p^0_2+M+16 q_{02}{c^1_0}^2x^2=0,
\end{equation}
where the coefficients in the polynomial $M$ are independent of
$q_{02}$. 

We differentiate relation (3.32) with respect to~$t$ and calculate
the derivatives $\dot p^0_1$ and $\dot p^0_2$ according to (3.28),
(3.34), (3.35), and (3.42). After elementary calculations, we obtain 
\begin{equation}%57
(\dot \sigma-7\mu\sigma \Delta p^{0}_2)\nabla p^{0\perp}_1
+(32 {c^1_0}^2q_{02}-14\mu\sigma^2)\nabla p^0_1=\cF^1.
\end{equation}
Here $\cF^1=\mu^{-1}D\dot{\oom}_2-\cM\nabla p^0_1$ and $\cM$
is the matrix with coefficients 
$\cM_{11}=7M_{20}-5M_{02}$, $\cM_{22}=7M_{02}-5M_{20}$, and 
$\cM_{12}=\cM_{21}=6M_{11}$, where $M_\alpha$ are 
coefficients in the polynomial~$M$. 

Obviously, (3.43) implies the differential equation
\begin{equation}%58
\dot \sigma-7\mu\sigma \Delta p^0_2
=\langle\cF^1,\nabla p^{0\perp}_1\rangle/{c^1_0}^2
\end{equation}
and the relation 
\begin{equation}%59
q_{02}=(7\mu\sigma^2+\langle\cF^1,\nabla p^0_1\rangle/2{c^1_0}^2)/
16{c^1_0}^2,
\end{equation}
which guarantee that (3.33) and (3.36) are compatible.

In turn, relations (3.33) and (3.45) mean that (3.42) can readily be
reduced to the scalar equation 
\begin{equation}%60
\dot c^0_{02}+7\mu\sigma^2+M_{02}+
\langle\cF^1,\nabla p^0_1\rangle/2{c^1_0}^2=0.
\end{equation}

Let us summarize the preceding. Relations (3.14) and (3.20) mean
that Eq.~(1.9) degenerates on the singularity support 
(since we have $c=c^0_0=1/2\mu$ there).
The construction of the first terms $p^0_1,p^0_2$ and 
$c^1_0,p^1_1$ in the asymptotic
expansion of the solution with respect to smoothness (3.1), (3.8) 
is reduced to solving ordinary differential equations (3.28),
(3.35), (3.41), (3.44), and (3.46) with regard to algebraic relations
(3.21), (3.27), (3.33), (3.40), and (3.45). 

We introduce the notation $y=(y_1,\dots,y_4)$, where 
$$
y_1=\sigma(t),\qquad
y_2=c^1_{10}(t),\qquad 
y_3=c^1_{01}(t),\qquad 
y_4=c^0_{02}(t),
$$
and $z=(z_1,\dots,z_7)$, where $z_j$ is a function 
from $c^0_{\alpha}(t)$, $|\alpha|=3$, or 
from $c^1_{\beta}(t)$, $|\beta|=2$.  
Then we can rewrite the above equations as follows:
\begin{eqnarray}%61
\dot a_1&=&-6\mu c^1_0\sin\varphi-\omega_{10},\nonumber\\
\dot a_2&=&-6\mu c^1_0\cos\varphi-\omega_{01},\nonumber\\
\dot \varphi&=&-7\mu y_1,\\
\dot c^1_0&=&\mu c^1_0\frac{9}{d'}(y_1\sin2\varphi-12y_4+2\omega_{20}/\mu),
\nonumber\\
\dot y_i&=&\cP_i(y,\omega_{20},\omega_{11}; \varphi,\mu)
+c^1_0L_i(\omega_{30},\omega_{03};\varphi,\mu)\nonumber\\
&&{}+M_i(\dot\omega_{20},\dot\omega_{11};\varphi,\mu)
+c^1_0 N_i(z;\varphi,\mu),\qquad i=1,\dots,4.\nonumber
\end{eqnarray}
Here $d'=\cos2\varphi-6$, 
$\cP_i(\tau;\varphi,\mu)$ are homogeneous second-order
polynomials in~$\tau$,  
$L_i(\xi;\varphi,\mu)$, $M_i(\xi;\varphi,\mu)$, and 
$N_i(z;\varphi,\mu)$ are homogeneous first-order polynomials  
in $\xi$ and~$z$.
The coefficients of these polynomials are uniformuly bounded
smooth functions of~$\varphi$ and the parameter
$\mu\geq\const>0$. 

The right-hand sides in (3.47) can readily be calculated 
by using computers and ``Mathematica'' software.
Nevertheless, the explicit formulas thus obtained 
are too cumbersome, and we do not write them here. 
We also note that system~(3.47)
is nonclosed, since it contains coefficients of the
polynomials $p^0_3$ and $p^1_2$.

Finally, 
it was proved that the function~$S$ is of the form 
$S=x^2(1+Q_1+Q_2+\dots)$, and
thus implies that the assumptions (3.2) are satisfied.

We consider the higher-order terms in the expansion with respect to
smoothness following the above scheme and arrive at the
statement of the theorem.

\begin{theor}
For the existence of a vortex type solution of Eq.~{\rm(1.9)},
it is necessary that the singularity support 
$x=-a(t)$ be the curve of degeneration of Eq.~{\rm(1.9)},
i.e., $c\big|_{x=-c(t)}=1/2\mu$,
and that the limit values of the derivatives of $c$
as $x\to-a(t)$ satisfy Hugoniot type conditions the first of
which have the form~{\rm(3.47)}. 
\end{theor}

For the initial data, it should be noted that 
relations (3.21) and (3.33) impose restrictions on the initial
values of $c^0_\alpha$, $|\alpha|=1,2$, 
while $c^1_\alpha\big|_{t=0}$, $|\alpha|\leq1$, 
and $\om_\alpha(t)$, $t\geq0$, can be chosen arbitrarily.

\begin{figure}
\hbox{\hbox to 0.5cm{}
\epsfxsize=160bp
\epsfbox[-74 164 549 778]{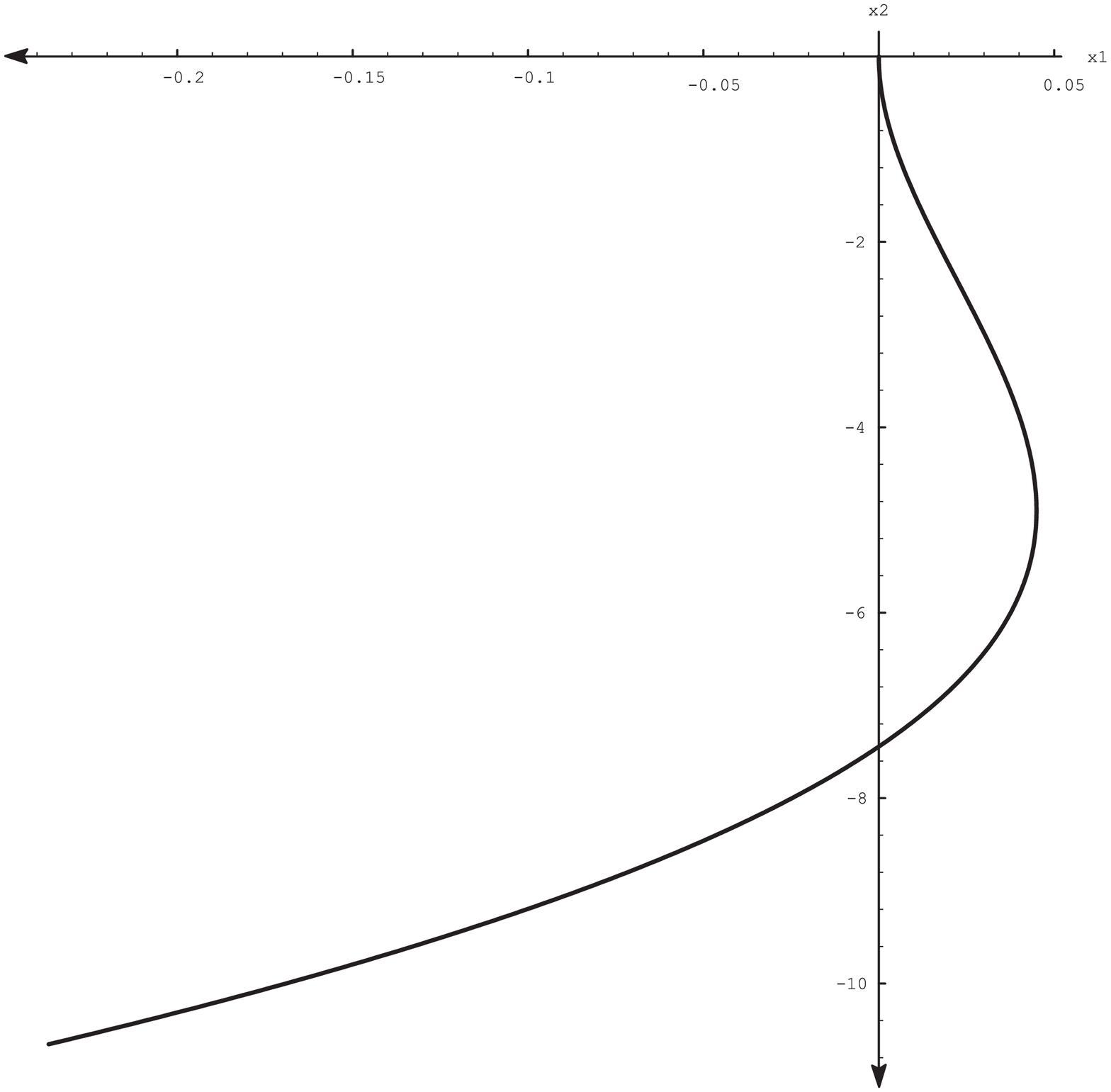}
\hbox to 1cm{}
\epsfxsize=160bp
\epsfbox[0 90 612 694]{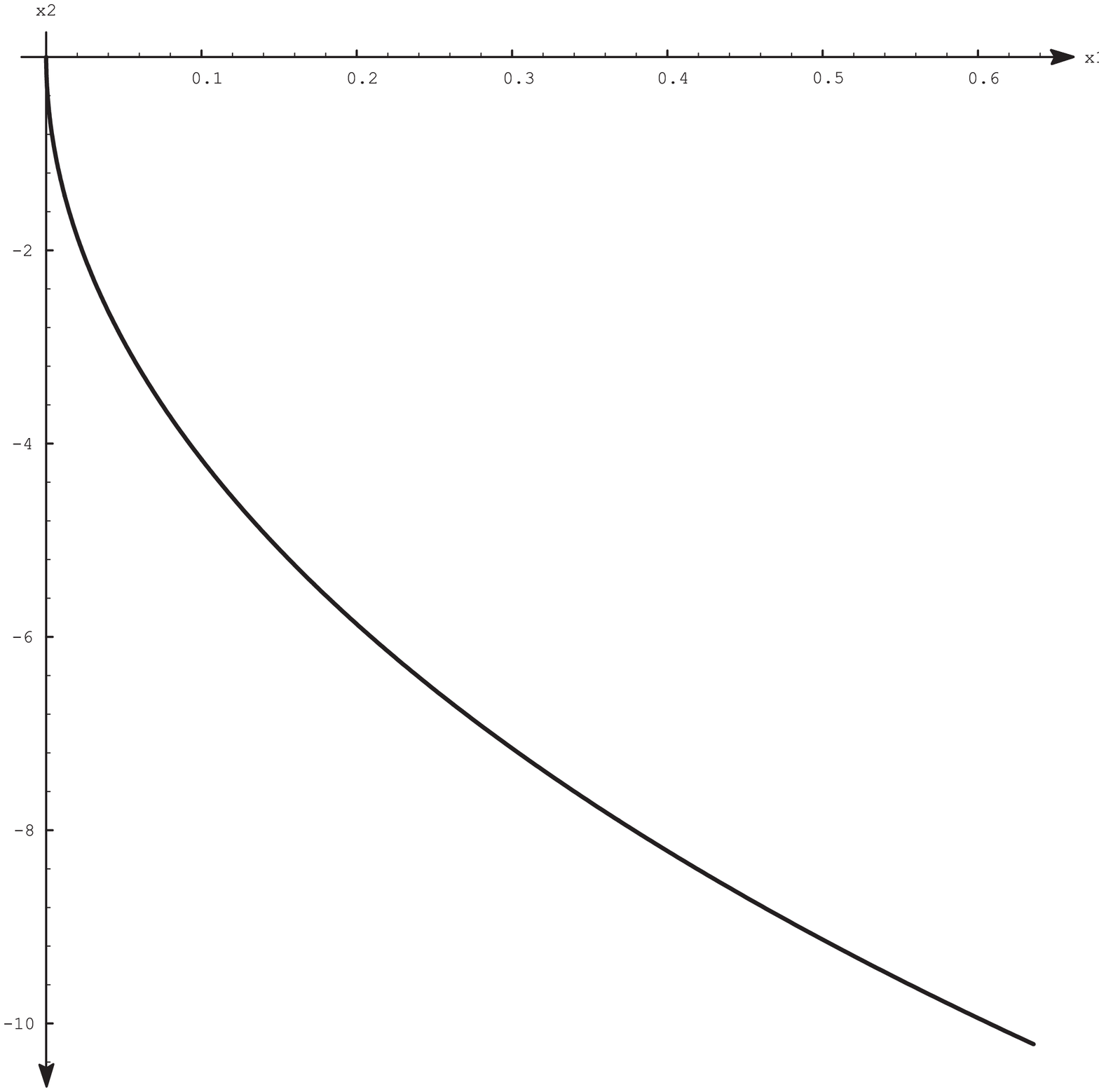}
}
\end{figure}

The plots show the trajectories 
(corresponding to different initial data)
of the singularity support motion calculated without regard for
the drift (for $\omega_\alpha=0$). 
These plots were calculated by using the truncated system (3.47),
where all ``extra'' functions $c^0_\alpha$, $|\alpha|=3$, 
and $c^1_\beta$, $|\beta|=2$, were set to be equal to zero.

The author is grateful to A.~I.~Koshelev and V.~A.~Galaktionov
who stimulated his interest in this problem.

This work was partially supported by the Russian Foundation 
for Basic Research under grants 99-01-01074 and 99-01-39128.


\begin{thebibliography}{99}
{\small
\bibitem{1}
Aronson D.~G., 
{\em Regularity properties of flows through porous media\rm: \em
the interface}, Arch. Rat. Mech. Anal., {\bf 37} (1970), 1--10. 
\bibitem{2} 
Biagioni H.~A. and Oberguggenberger M.,
{\em Generalized solutions to Burgers' equation}, 
J. Diff. Eq., {\bf 97}:2 (1992), 263--287.
\bibitem{3} 
Biagioni H.~A. and Oberguggenberger M.,
{\em Generalized solutions to the Korteweg-de Vries and the
regularized long-wave equations}, 
SIAM J. Math. Anal., {\bf 23}:4 (1992), 923--940.
\bibitem{4}
Bulatov V.~V., Vladimirov Yu.~V., Danilov V.~G., and 
Dobrokhotov S.~Yu.,  
{\em Propagation of a pointwise algebraic singularity for
two-dimensional nonlinear equations of hydrodynamics}, 
{\em Mat. Zametki}, {\bf55}:3 (1994), 11--20 (in Russian);
English translation in Math. Notes.
\bibitem{5}
Caffarelli L. and Friedman A.,
{\em Regularity of the free boundary of a gas flow in an
$n$-dimensional porous medium}, Indiana Univ. Math. J., 
{\bf 29} (1980), 361--391.
\bibitem{6}
Cauret J.~J., Colombeau J.~F., and Le Roux A.~Y.,
{\em Discontinuous generalized solutions of nonlinear
nonconservative hyperbolic equations}, J. Math. Anal. Appl.,
{\bf 139} (1989), 552--573. 
\bibitem{7}
Danilov V.~G., Maslov V.~P.,  and Volosov K.~A., 
{\em Mathematical Modelling of Heat and Mass Transfer
Processes}, Kluwer, Dordrecht,  1995.
\bibitem{8}
Danilov V.~G. and Omel'yanov G.~A.,
{\em Truncation of a chain of Hugoniot-type conditions 
for  shock waves and  its justification for the Hopf equation},
Preprint ESI No.~502, 1997.
\bibitem{9}
Danilov V.~G., Maslov V.~P.,  and Shelkovich V.~M., 
{\em Algebras of singularities of singular solutions to
quasilinear strongly hyperbolic first-order equations},   
Theoret. and Math. Phys., {\bf 114} (1998), 3--55.
\bibitem{10}
Di Benedetto E.,
{\em Regularity results for the porous media equation},
Ann. Mat. Pura Appl., {\bf 121} (1979), 249--262.
\bibitem{11} 
Djapi\'c N., Pilipovo\'c S., and Scarpal\'ezos D., 
{\em Microlocal analysis of Co\-lom\-beau's generalized functions 
-- propagation of singularities}, J. D'Anal. Math., 75 (1998), 51--66.
\bibitem{12} 
Djapi\'c N. and Pilipovo\'c S., 
{\em Approximated traveling wave solutions to generalized Hopf
equation}, Novi Sad J. Math., 1998.
\bibitem{13}
Friedman A., 
{\em Variational Principles and Free Boundary Problems}, 
Wiley, New York, 1982.
\bibitem{14} 
Galaktionov V.~A., Hulshof J., and Vazquez J.~L.,
{\em Extinction and focusing behavior of spherical and annular
flames described by a free boundary problem}, 
J. Math. Pure Appl., {\bf 76} (1997), 563--608.
\bibitem{15}
Gilding B.~H.,
{\em Properties of solutions of an equation in the theory of
infiltration}, Arch. Rat. Mech. Anal., {\bf 65} (1977), 203--225. 
\bibitem{16}
Kershner R.,
{\em Several properties of generalized solutions of quasilinear
degenerate parabolic equations}, 
Acta Math. Acad. Sci. Hungaricae, {\bf 32}:3--4 (1978),
301--330. 
\bibitem{17}
Knerr B.~F.,
{\em The behavior of the support of solutions of the equations
of nonlinear heat conduction with absorption in one dimension},
Trans. Amer. Math. Soc., {\bf 249} (1979), 409--424.
\bibitem{18}
Maslov V.~P., 
{\em Three algebras corresponding to nonsmooth solutions of
systems with quasilinear hyperbolic equations}, 
Russ. Math. Surv., {\bf 35}:2 (1980), 252--253.
\bibitem{19}
Maslov V.~P.,   {\em Propagation of shock waves in an isoentropic
nonviscous gas}, Itogi Nauki i Tekhniki, 8, 
VINITI, Moscow, 1977 (In Russian.)
\bibitem{20} 
Meirmanov A.~M.,
{\em The Stefan Problem}, Nauka, Novosibirsk, 1986 (in Russian).
\bibitem{21} 
Nedeljkov M.,
{\em Infinitely narrow soliton solutions to scalar
conservation laws in Coloumbeau sense},  Integral
Transformations and Special Functions, 6 (1998), 1--4, 257--263.
\bibitem{22}
Peletier L.~A.,
{\em The Porous Media Equation. 
Applications of Nonlinear Analysis in the Physical Sciences}, 
Pitman, Boston--Melbourne, 1981, 229--241.
\bibitem{23}
Ravindran R. and Prasad P.,  {\em A new theory of
shock dynamics},    Appl. Math. Lett., 
{\bf 3}:2 (1990), 107--109.
\bibitem{24}
Samarskii A.~A., Galaktionov V.~A., Kurdyumov S.~P., and
Mikhailov A.~P., 
{\em Blow-up in Quasilinear Parabolic Equations}, Nauka, Moscow,
1987 (in Russian); English translation: Walter de Gruyter,
Vol.~19, Berlin--New York, 1995.
\bibitem{25} 
Vazquez J.~L.,
{\em The free boundary problem for the heat equation with fixed
gradient condition}, Proceedings of the Conference ``Free
Boundary Problems: Theory and Applications'', Zakopane, Poland,
June 1995.

}
\end{thebibliography}
\end{document}